\documentclass[%
 reprint,
showpacs,preprintnumbers,
nofootinbib,
 amsmath,amssymb,
 aps, 
]{revtex4-1}

\input psfig.sty

\usepackage{graphicx}
\usepackage{dcolumn}
\usepackage{bm}

\begin{document}

\preprint{GCON/08-2015}

\title{A low-$z$ test for interacting dark energy}

\author{R. S. Gon\c{c}alves}
 \email{rsousa@on.br}
 \author{G. C. Carvalho}
 \email{gabriela@on.br}
\author{J. S. Alcaniz}
 \email{alcaniz@on.br}
\affiliation{Observat\'orio Nacional, 20921-400, Rio de Janeiro, RJ, Brazil }

\date{\today}

\begin{abstract}

A non-minimal coupling between the dark matter and dark energy components may offer a way of solving the so-called coincidence problem. In this paper we propose a low-$z$ test for such hypothesis using measurements of the gas mass fraction $f_{\rm{gas}}$ in relaxed and massive galaxy clusters. The test applies to any model whose dilution of dark matter is modified with respect to the standard $a^{-3}$ scaling, as usual in interacting models, where $a$ is the cosmological scale factor. We apply the test to current $f_{\rm{gas}}$ data and perform Monte Carlo simulations to forecast the necessary improvements in number and accuracy of upcoming observations to detect a possible interaction in the cosmological dark sector. Our results show that  improvements in the present relative error $\sigma_{\rm{gas}}/f_{\rm{gas}}$ are more effective to achieve this goal than an increase in the size of the  $f_{\rm{gas}}$ sample.

\end{abstract}


\maketitle

\section{Introduction}

One of the most important goals of current cosmological studies is to unveil the nature of the mechanism behind cosmic acceleration. In principle, the most promising explanations involve either the existence of new fields in high energy physics or large-scale modifications of gravity theory. Following the former route, several possibilities have been discussed in the literature, among them a possible interaction between the dark matter and the smooth dark energy component, permitting transfer of energy between these two fields and violating adiabaticity (see \cite{ltcdm} and references therein).

If the dark energy component is the vacuum energy density $\rho_{\Lambda}$, characterised by an equation-of-state parameter $w=p/\rho=-1$, where $p$ is the pressure, these scenarios may constitute not only a phenomenological attempt at alleviating the so-called coincidence problem, since dark matter and dark energy are no longer assumed to be independent, but also the unsettled situation in the particle physics/cosmology interface, in which the cosmological upper bound for $\rho_{\Lambda}$ differs from theoretical expectations ($\rho_{v} \sim 10^{71} {\rm{GeV}}^4$) by more than 100 orders of magnitude (see \cite{weinberg} for a discussion on the cosmological constant problems). From the physical viewpoint, it worth mentioning that no known symmetry in Nature prevents a non-minimal coupling between these components. Therefore, although no observational piece of evidence has so far been unambiguously presented for an interaction in the cosmological dark sector, a weak coupling still below current detection 
cannot be 
completely excluded.

Given the theoretical implications of such a detection for cosmology, it is important to investigate ways to probe this possibility from observations. Our goal in this paper is to propose and explore a low-$z$ test for a possible dark matter/dark energy interaction from measurements of the gas mass fraction $f_{\rm{gas}}$ in relaxed and massive galaxy clusters.  The test proposed is model-independent in the sense that in the low-$z$ regime the $f_{\rm{gas}}$ equations become independent of the interacting model, although the signature of a non-minimally coupled remains present. We apply the test to the current data and investigate the constraining power of future $f_{\rm{gas}}$ measurements from Monte Carlo (MC) simulations. We also discuss the necessary conditions in terms of number and accuracy of upcoming data for a successful application of the test.  In what follows, we outline the main assumptions of our analysis and discuss the main results.

\section{Interaction in the dark sector}

In ou analysis, we assume a non-minimal coupling between the cold dark matter (CDM) and a dark energy component ($\Lambda$) with $w=-1$ (baryons and radiation are assumed to be separately conserved, implying the usual scaling $\rho_b \propto a^{-3}$ and $\rho_{\gamma} \propto a^{-4}$). The evolution of the mean dark matter density $\rho_m$ and of $\rho_{\Lambda}$ are obtained from the balance equations
\begin{equation} \label{c1}
\dot \rho_{\Lambda} = - {\cal{Q}}\;,
\end{equation}
\begin{equation} \label{c2}
\dot{\rho}_{m} + 3H{\rho}_{m} = -\cal{Q}\; ,
\end{equation}
where a dot denotes derivative with respect to time, $H = \frac{\dot{a}}{a}$ is the Hubble parameter and $a$ is the cosmological scale factor, written in terms of the redshift parameter as $a = 1/(1+z)$. In the absence of a natural guidance from fundamental physics on the interaction function $\cal{Q}$, which  controls the rate of energy transfer, several phenomenological forms have been proposed~\cite{ltcdm}. An interesting approach was discussed in Ref.~\cite{wm} in which the time variation of $\Lambda$ is deduced from its effect on the CDM evolution. The qualitative argument is the following: since $\Lambda$ is decaying into CDM particles, CDM will necessarily dilute more slowly compared to its standard evolution, $\rho_m \propto a^{-3}$. Thus, if the deviation from the standard evolution is characterised by a constant $\epsilon$, i.e.,
\begin{equation} \label{eq1}
\rho_m \propto \varphi(a)a^{-3}\, \quad \mbox{where} \quad \varphi(a) = a^{\epsilon}\;,
\end{equation}
Eqs.(\ref{c1}) and (\ref{c2}) yield 
\begin{equation} \label{lambda}
{\rho}_{\Lambda} =  \frac{\epsilon {\rho}_{m,0}}{3 - \epsilon} a^{-3+\epsilon}  + \mbox{const.}\; \; ,
\end{equation}
which is equivalent to the dark energy evolution found if one assumes ${\cal{Q}} = \epsilon \rho_m H$ (from now on, the subscript 0 denotes present-day quantities). Note that a positive (negative) value of $\epsilon$ leads to a slower (faster) dilution of the matter component, which implies in an energy transfer from $\Lambda$ (CDM) to CDM ($\Lambda$) (for thermodynamic constraints on the sign of $\epsilon$, we refer the reader to \cite{Alcaniz05,Pavon}). A more general parameterisation for the interaction parameter was discussed in Ref.~\cite{costa}, in which this quantity is assumed to be an arbitrary function of the scale factor, i.e., $\epsilon(a)$. In this case, regardless of the interaction function $\cal{Q}$ a given theory may provide or, equivalently, of its effect on the CDM evolution, $\rho_m$ can always be parameterised in terms of $\epsilon(a) = \log_a\varphi(a)$. Note also that this approach can be applied even if the dark energy component has an arbitrary equation-of-state parameter ($w(z) \neq 
-1$).

\begin{figure*}
\includegraphics[width=2.3in,height=2.7in]{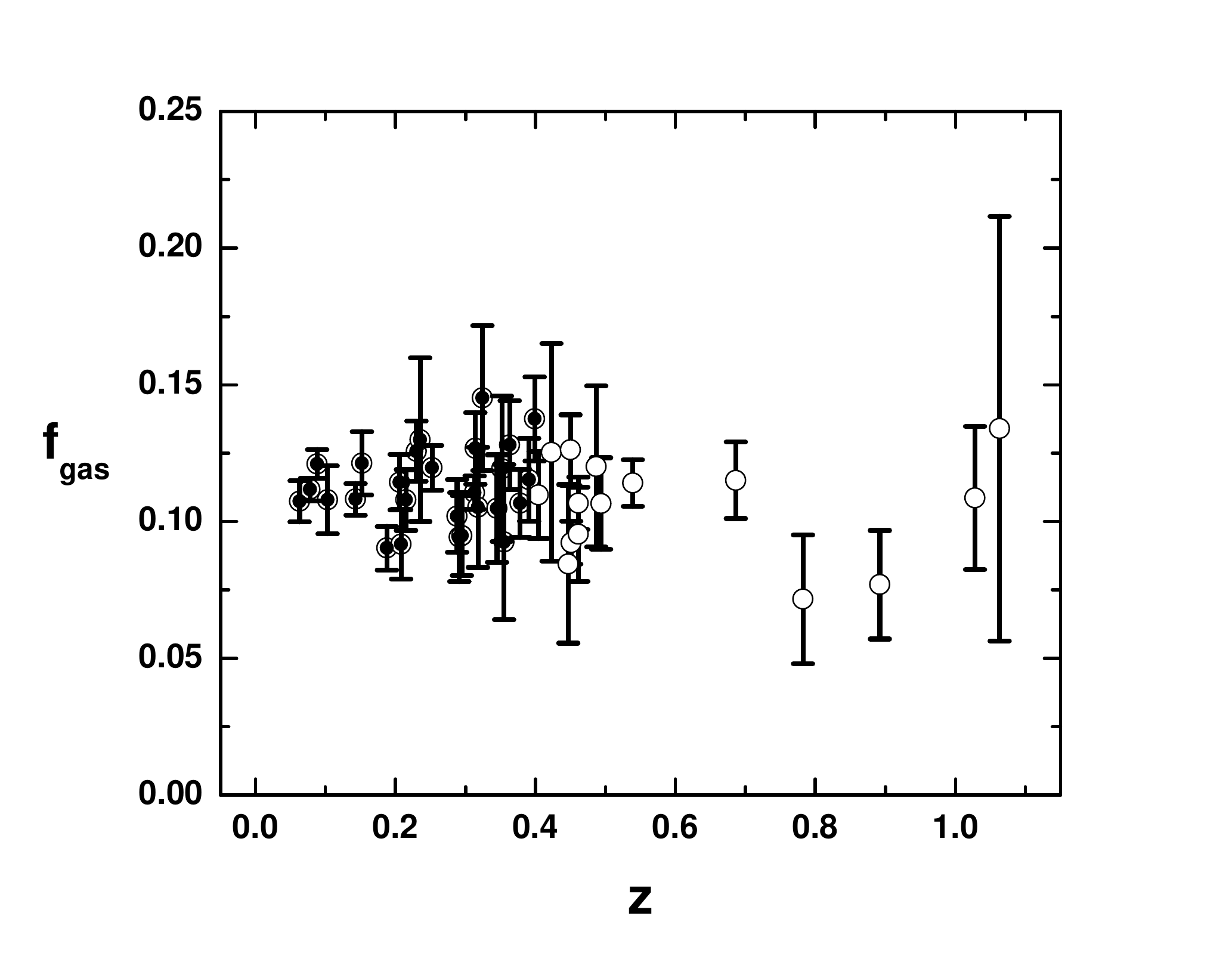}
\includegraphics[width=2.3in,height=2.7in]{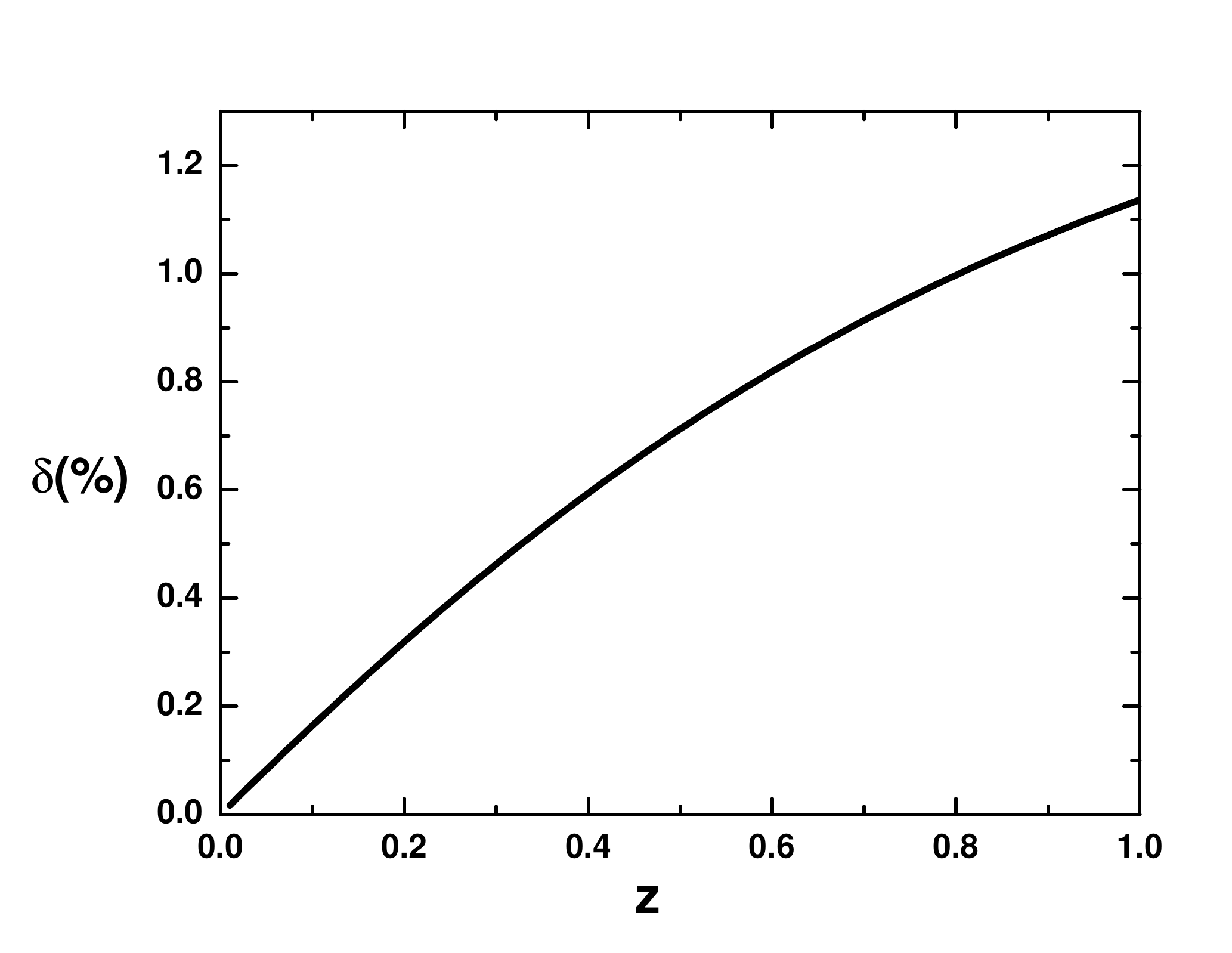}
\includegraphics[width=2.3in,height=2.7in]{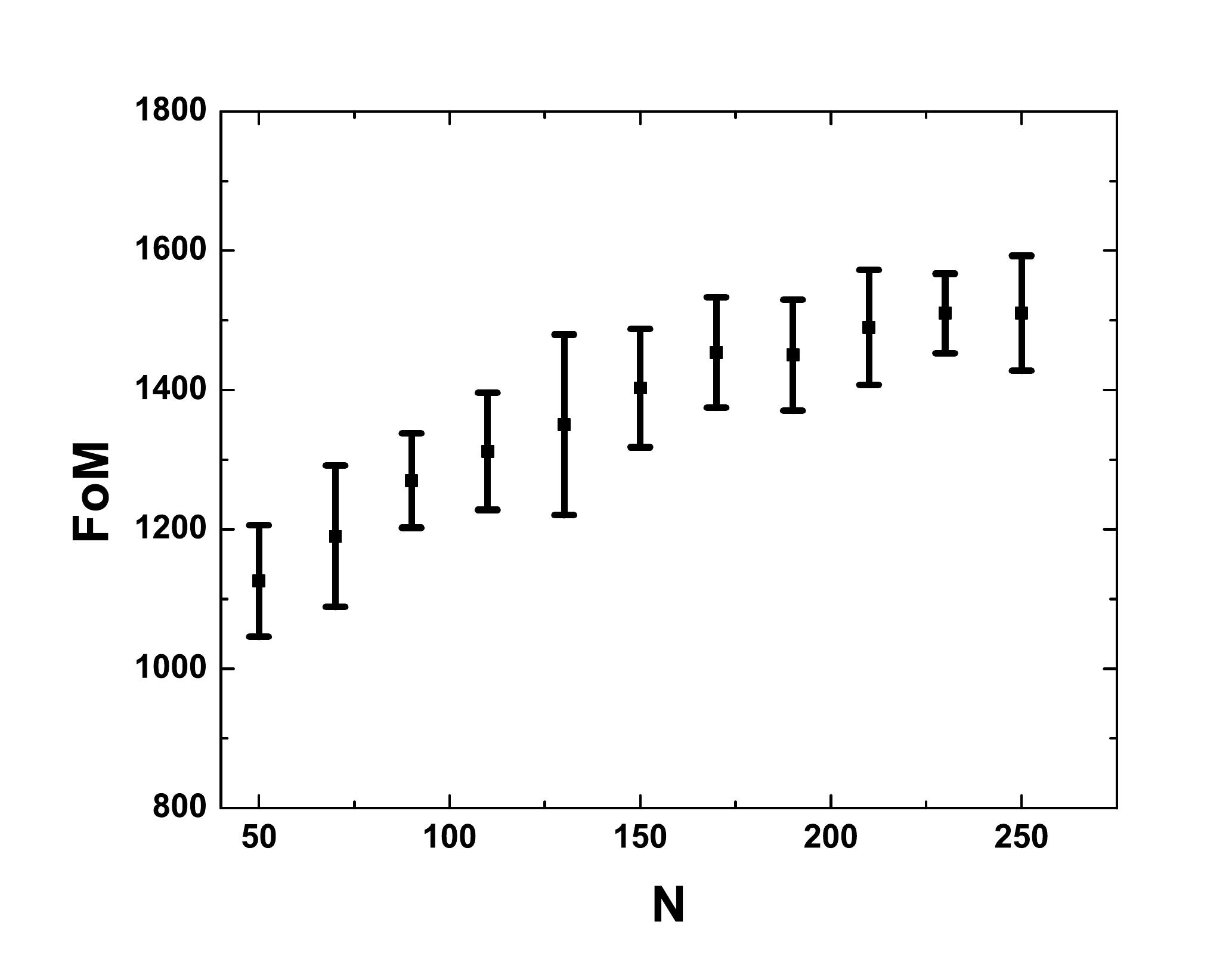}
\caption{ {\emph{Left)}} Gas mass fraction measurements as a function of redshift for 42 galaxy clusters studied in Ref. \cite{allen08}. Solid circles corresponds to the reduced sample (27 data  points) obtained after applying the cosmographic condition discussed in the text. {\emph{Middle)}} Dispersion of the angular diameter distance  as function of the redshift. The maximum limit allowed is  $\delta = 0.6\%$, which reduces our sample to clusters up to $z \leq 0.4$. {\emph{Right)}} Evolution of the figure of merit defined in the text as a function of the number of data points $N$. The current value, obtained from the 27 data points shown in the panel of the left, is FoM$_{obs} \simeq 425$} 
\label{fig1}
\end{figure*}

\section{A low-$z$ test for interacting dark energy}

Observations of the gas mass fraction in relaxed and massive galaxy clusters have been widely used as a cosmological test (see, e.g., \cite{fgas,allen02,lca,lr,allen08,ettori,allen13} and references therein). The key assumption of this test is that the baryonic mass fraction in galaxy clusters provides a fair sample of the distribution of baryons at large scale \cite{White93}. The mass fraction of hot gas measured within a characteristic radius of a halo at redshift $z$ can be written as
\begin{equation}
\label{Fgas}
f_{gas} (z)= {\cal{G}} \frac{\Omega_{b}(z)}{\Omega_{b}(z) + \Omega_{m}(z)}\left[\frac{d^{fiducial}_{A}(z)}{d^{model}_{A}(z)} \right]^{\frac{3}{2}}\;,
\end{equation}
where $\Omega_b(z)$ and $\Omega_m(z)$ are, respectively, the baryon and CDM density parameters as function of $z$ and 
the angular diameter distance ($d_A$) ratio accounts for deviations in the geometry of the universe between the fiducial model adopted in the observations and the model to be tested\footnote{For a general derivation of $f_{gas} (z)$ taking into account possible violations of the so-called distance duality relation, see \cite{Goncalves12}.}. The term $\cal{G}$ accounts for star formation and other baryon effects within the characteristic radius under the assumption that $f_{gas}$ should be approximately constant with redshift (for a discussion of the systematics errors included on $\cal{G}$, see also \cite{allen13}). Many forms of $\cal{G}$ have been adopted in the literature (see, e.g., \cite{allen02,allen08,ettori,allen13}). However, for the test discussed here the exact form or value of $\cal{G}$ is not important, so that we marginalise over it in all the analyses presented in the next sections.

Regardless of the cosmological models involved in the analysis, the distance ratio above equals unity in the cosmographic limit ($z << 1$). Moreover, if $\epsilon = 0$, the baryon-to-total-mass density ratio ${\Omega_{b}(z)}/[{\Omega_{b}(z) + \Omega_{m}(z)}]$ becomes constant, as the baryons and CDM density parameters have the same $z$ dependence. On the other hand, under the assumptions discussed earlier, if there is a non-minimal coupling in the cosmological dark sector the above equation can be rewritten as 
\begin{equation}
\label{FgasLowZ}
f_{gas} (z<<1)= {\cal{G}} \frac{\Omega_{b,0}}{\Omega_{b,0} + \Omega_{m,0}(1+z)^{-\epsilon(z)}} \;,
\end{equation}
where the extra term $(1+z)^{-\epsilon(z)}$ makes $f_{gas} (z<<1)$ no longer constant, providing a direct test to the idea of an interacting dark energy field. In reality, this test can be applied to any scenario whose dilution of dark matter is modified with respect to the usual $a^{-3}$ scaling, regardless whether it is due to an interacting or any other physical process. Here, however, for simplicity, we will restrict our analysis to the case in which a non-minimal coupling in the dark sector is parameterised by a constant interacting parameter $\epsilon$, as given in Eq.(\ref{eq1}). 

\section{Implementation of the test}

\subsection{Data}

In our analysis, we consider the Chandra $f_{gas}$ data analysed in \cite{allen08} and shown in Fig. 1a. Originally, this sample comprises $42$ regular, relatively relaxed and massive galaxy clusters distributed over a wide range of redshift ($0.063 < z < 1.063$). 

\subsection{Current constraints}

In order to implement the low-$z$ test of Eq. (\ref{FgasLowZ}), we first study the redshift interval in which the cosmographic condition above is satisfied. The fiducial model used in the observations was a flat $\Lambda$CDM scenario with $\Omega_{m,0} = 0.3$ \cite{allen08}. Thus, considering $\Omega_b = 0.0489$, given by current CMB experiments \cite{planck}, we show in figure 1b the quantity $\delta = [(d_A^{\epsilon \rm{CDM}}-d_A^{\Lambda \rm{CDM}})/d_A^{\Lambda \rm{CDM}}] \times 100\%$, where $\epsilon$CDM stands for the interacting model discussed in Sec. II with the values of $\epsilon$ and $\Omega_{m,0}$ fixed at the  best-fit values given by Eq. (\ref{r1}). We consider $f_{gas}$ data up to $z = 0.4$, which corresponds to a  dispersion of $0.6\%$ for the angular diameter ratio of Eq. (\ref{Fgas}). Such value is negligible when compared with current observational uncertainties ($\sigma_{f_{gas}}/f_{gas} \simeq 11.4\%$) and restricts our working sample to $27$ clusters (Fig. 1a)\footnote{Note that this 
large $z$ interval is possible only because the fiducial model used in the $f_{gas}$ observations and the best-fit scenario found below [Eq. (\ref{r1})] predict very similar $d_A(z)$ evolution. However, the most general way to define the test of Eq. (\ref{FgasLowZ}) is through the cosmographic limit discussed above.}.

Considering this sub-sample, we perform a $\chi^2$ analysis: 
\begin{equation}
\label{Chi2Geral}
\chi^2 = \sum_{i = 1}^{27} \frac{\left[ f_{gas}^{obs}(z_i) - f_{gas}^{th}(z_i;\epsilon,\Omega_{m,0}) \right]^{2}}{\sigma^2_{f^{obs}_{gas}} }\;,
\end{equation}
where $f_{gas}^{th}$ is the theoretical expression for $f_{gas}$ at low-$z$ given by Eq. (\ref{FgasLowZ}) and $\sigma^2_{f^{obs}_{gas}}$ is the error in the $f_{gas}^{obs}$ measurements. 
From this analysis, we find
\begin{equation} \label{r1}
\Omega_{m,0} = 0.28 \pm 0.01 \quad \mbox{and} \quad \epsilon = 0.035^{+0.215}_{-0.235} \quad (1\sigma)\;,
\end{equation}
with $\chi^2_{\nu} = 1.17$ ($\nu$ stands for the number of degrees of freedom). Clearly, the current data are unable to unambiguously detect or rule out a non-minimal coupling in the cosmological dark sector. In what follows, we perform MC simulations to discuss the necessary conditions (i.e., number of $f_{gas}$ data points and improvements on the relative error measurement $\sigma_{f_{gas}}/f_{gas}$) to detect such a coupling at a given confidence level. 

\begin{figure*}
\includegraphics[width = 3.25in, height = 3.4in]{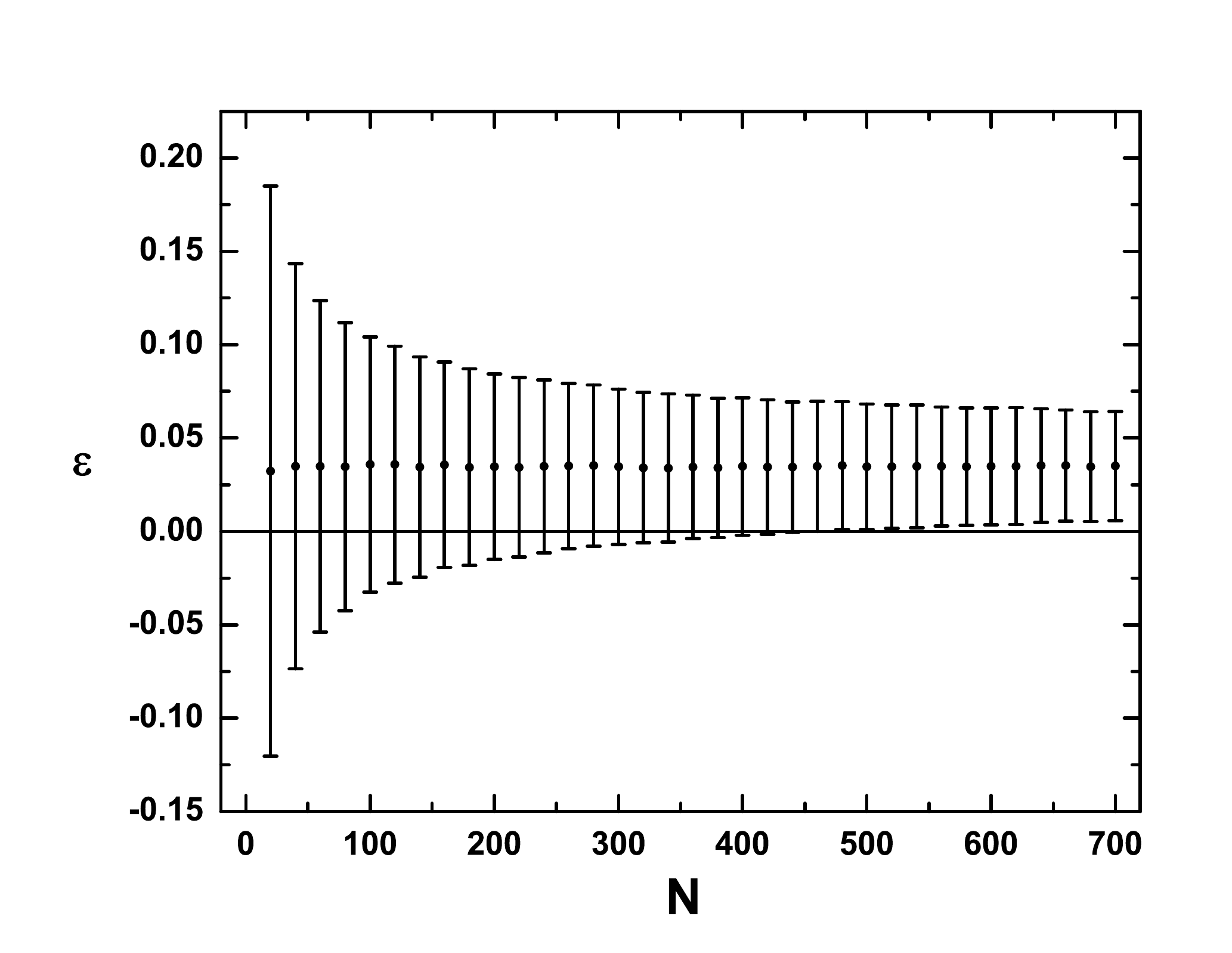}
\hspace{0.5cm}
\includegraphics[width = 3.25in, height = 3.4in]{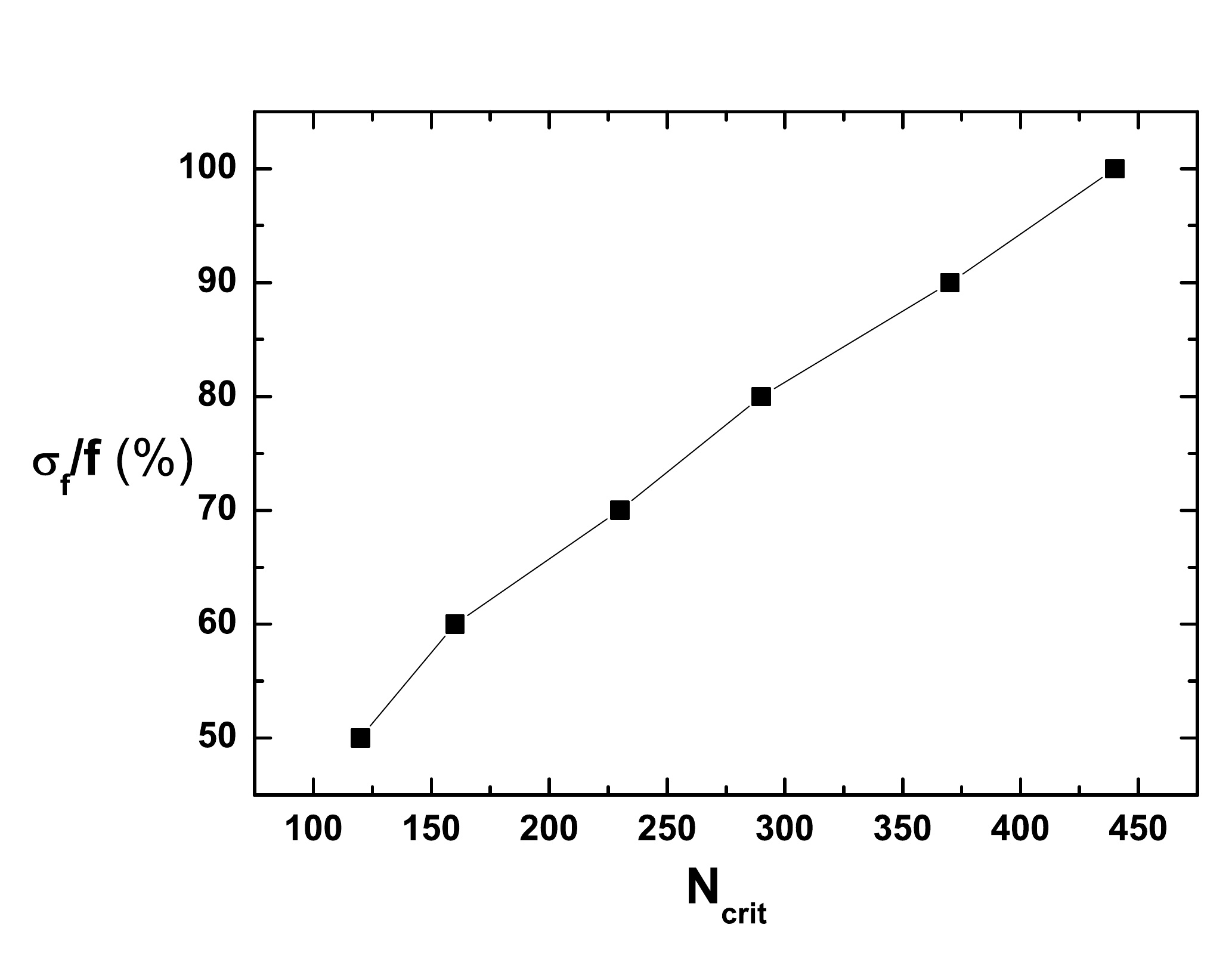}
\caption{ {\emph{Left}} The interacting parameter $\epsilon$ as a function of the number of data points $N$. Assuming the values given in Eq. (\ref{r1}) as fiducial value, the number of points for which $\sigma_{\epsilon} = \epsilon$ is $N_{crit} = 440$. {\emph{Right}} Percentage of the relative error as a function of the critical number of data points. For $\sigma_{f_{gas}}/f_{gas} = 100\%$, the results of the previous panel are recovered.}
\end{figure*} 

\section{Simulations and results}

We expect that the larger the number of $f_{gas}$ data points or the smaller their error $\sigma_{f_{gas}}$, the smaller the uncertainty on the interacting parameter $\sigma_{\epsilon}$ should be. Thus, in order to find which are the necessary improvements in future observations to check the hypothesis of a non-minimal coupling in the cosmological dark sector, we generate synthetic samples of $f_{gas}$ measurements assuming the distribution and redshift interval of the current observations.

The steps of the simulations are the following: we first define the number of points ($N$) to be simulated. Then, we simulate a sample of $f_{gas}(z)$ measurements, with $N$ points in the same redshift range of the current data. For each point simulated we assume a normal distribution ${\cal{N}}(\mu,\sigma)$, with a mean value and error defined as the fiducial values. The relative errors for $f_{gas}$ are obtained assuming that the relative error provided by observations can be fitted by a polynomial to find its redshift dependence. We performed $10^4$ realisations for each one of the analyses described below.

We first study the evolution of the Figure-of-Merit, defined as FoM = $1/(\sigma_{\epsilon} \sigma_{\Omega_{m,0}})$, as function of the number of data points in the range [50 - 250]. Using the simulated data we perform $\chi^2$ analyses and find the best-fit values for the pair ($\epsilon$, $\Omega_{m,0}$). We repeat this process $10^4$ times and calculate the mean and standard deviation of these parameters and  the corresponding FoM, for a fixed value of $N$. The fiducial model used in the simulations has the parameters values given in Eq. (\ref{r1}) and the results are shown in Fig. 1c. From this analysis, we find that for a relatively small number of measurements when compared to future X-ray surveys\footnote{See, e.g., http://www.mpe.mpg.de/eROSITA}, e.g., $N = 200$, the FoM value $~1.5 \times 10^3$ corresponds to an improvement of order of 3 with respect to the current data (FoM$_{obs} \simeq 425$). This clearly shows that very tight bounds on the interacting parameter can be obtained from a larger 
sample of galaxy clusters X-ray observations.

In what follows, we will direct our analysis to the necessary improvements in future observations, i.e., the number of galaxy clusters $N$ and the relative error of the gas mass fraction $\sigma_{f_{gas}}/f_{gas}$, to observationally distinguish between values of $\epsilon$ of the order of current estimates and the standard value $\epsilon = 0$. Besides assuming the best-fit values given in Eq. (\ref{r1}) as fiducial values, our simulations always assume the distribution and relative error of the current $f_{gas}$ sample shown in Fig. 1a. 

Under such conditions, the simulation process provides the number of data points $N_{crit}$ necessary to verify the hypothesis of a non-minimal coupling in the cosmological dark sector.  In order to find $N_{crit}$, we adopt the criterion $n \sigma_{\epsilon} /\epsilon < 1$, where $n$ corresponds to the confidence level (number of standard deviations) required. The results are shown in the Fig. 2a. As expected, the error bar for the interaction parameter decreases for higher values of $N$. For $n = 1$ ($1\sigma$), we find $N_{crit} = 440$, which clearly illustrates the viability of implementation of the test procedure in the near future since a value of $N >> N_{crit}$ is expected in some planned X-ray surveys. It is worth noticing that such result is somehow conservative since no improvement on the upcoming $f_{gas}$  data relative to the current ones was considered to derive the above value of $N_{crit}$. 

In order to study the influence of the uncertainties of $f_{gas}$ measurements on the estimates of $N_{crit}$, we also vary the relative error $\sigma_{f_{gas}}/f_{gas}$ with $N$, considering the interval between $50\%-100\%$ of the current observational value. This is shown in Fig. 2b, where 100\% means the present-day value. Clearly, by improving the quality of the data or, equivalently, decreasing their relative error, $N_{crit}$ becomes  considerably smaller. As an example, decreasing the relative error  in 50\%, we find $N_{crit} = 145$, which corresponds to 1/3 of the previous value obtained when the relative error of the current $f_{gas}$ data ($\sigma_{f_{gas}}/f_{gas} = 100\%$) is considered. For $N_{crit} = 27$, which corresponds to the number of $f_{gas}$ measurements currently available, the improvement on  $\sigma_{f_{gas}}/f_{gas}$  needed to distinguish coupled from uncoupled dark energy is of the order of 25\% relative to the present observational value. These results indicate that 
improvements in the relative error of $f_{gas}$ observations are more effective to achieve this goal than an increase in the size of the sample.

\section{Concluding remarks}

Testing a possible non-minimal coupling between the dark matter and dark energy fields constitutes an important task for cosmology and fundamental physics. In this paper, we have proposed and applied a new, model-independent test for this hypothesis using low-$z$ measurements of the gas mass fraction in galaxy clusters. As well known, the transfer of energy between the dark components necessarily modifies the dilution of dark matter with respect to the usual $a^{-3}$ scaling whereas the baryonic component remains separately conserved ($\rho_b \propto a^{-3}$). Therefore, the test here proposed can in principle be applied to any model of this class or even to cosmologies in which the attenuated dilution of the CDM component results from another physical process. This is possible because in the low-$z$ regime the $f_{\rm{gas}}$ equations become cosmological model-independent, although the signature of a non-standard CDM evolution remains present. 

Since the current observational data cannot unambiguously detect any interaction between the dark components, we have used the present bounds given by Eq. (\ref{r1}) as fiducial values and performed MC simulations to study the necessary requirements for a $\epsilon \neq 0$ detection. We have then generated thousands of $f_{\rm{gas}}$ samples assuming the present observational error distribution, as described in Sec. V.  After discussing the gain in the FoM with the number of data points $N$ in the synthetic sample, we have investigated the number and accuracy of future observations needed for a successful application of the test.  For the simplest case in which $\epsilon$ = const., adopted in our analysis, and considering the relative error $\sigma_{f_{gas}}/f_{gas}$ of current data, we have found that the minimum number of data points necessary to distinguish between an interacting and a non-interacting dark energy component should be $\sim 15$ times larger than the current observational sample ($N_{crit} = 
440$ and $N_{obs} = 27$). This number, however, is much smaller than the number of galaxy clusters expected to be observed by the extended Roentgen Survey with an Imaging Telescope Array (eROSITA), which clearly shows the viability of implementation of this $f_{gas}$ test in the near future. We have also calculated how $N_{crit}$ changes as the relative error of $f_{gas}$ measurements improves. As shown in Fig. 2b, decreasing $\sigma_{f_{gas}}/f_{gas}$ by a factor of 2 leads to $N_{crit} = 145$, which is three times smaller than the value found assuming the relative error of the current $f_{gas}$ data. These results clearly show that a combination of quantity and precision of future $f_{gas}$ measurements can become an important tool not only to probe a possible interaction in the cosmological dark sector but also to explore its consequences.

\section*{Acknowledgements}

The authors thank CNPq, INEspa\c{c}o and FAPERJ for the grants under which this work was carried out. RSG is supported by the DTI-PCI fellowship program of the Brazilian Ministry of Science, Technology and Innovation (MCTI).

\end{document}